# Analysis of a Modern Voice Morphing Approach using Gaussian Mixture Models for Laryngectomees

Aman Chadha
Department of Electrical and
Computer Engineering
University of Wisconsin-
Madison

Bharatraaj Savardekar
Department of Electronics
Engineering
K. J. Somaiya College of
Engineering

Jay Padhya
Department of Electronics
Engineering
K. J. Somaiya College of
Engineering

## ABSTRACT

This paper proposes a voicemorphing system for people suffering from Laryngectomy, which is the surgical removal of all or part of the larynx or the voice box, particularly performed in cases of laryngeal cancer. A primitive method of achieving voice morphing is by extracting the source's vocal coefficients and then converting them into the target speaker's vocal parameters. In this paper, we deploy Gaussian Mixture Models (GMM) for mapping the coefficients from source to destination. However, the use of the traditional/conventional GMM-based mapping approach results in the problem of over-smoothening of the converted voice. Thus, we hereby propose a unique method to perform efficient voice morphing and conversion based on GMM, which overcomes the traditional-method effects of over-smoothening. It uses a technique of glottal waveform separation and prediction of excitations and hence the result shows that not only over-smoothening is eliminated but also the transformed vocal tract parameters match with the target. Moreover, the synthesized speech thus obtained is found to be of a sufficiently high quality. Thus, voice morphing based on a unique GMM approach has been proposed and also critically evaluated based on various subjective and objective evaluation parameters. Further, an application of voice morphing for Laryngectomees which deploys this unique approach has been recommended by this paper.

## General Terms

Audio and Speech Processing

## Keywords

Voice Morphing, Laryngectomy, Gaussian Mixture Models

## 1. INTRODUCTION

Voice morphing is essentially a technique of transforming a source speaker's speech into target speaker's speech. In general, the voice morphing systems consist of two stages: training and transforming. Out of the two, the main process is the transformation of the spectral envelope of the source speaker to match to that of the target speaker. In order to implement the personality transformation, there are two issues that need to be tackled: firstly, converting the vocal tract feature parameters as well as transformation of excitation parameters. Till recently, several published works [1]-[5] in the domain of voice conversion have been focusing on the vocal tract mapping whose features are parameterized by respective Linear Predictive Coding (LPC) parameters. However it has known that some kinds of transformation are needed to be applied to excitation signals in order to achieve transformations of high quality. Furthermore, the converted speech often suffers from the degraded quality due to over-smoothening effects caused by the traditional GMM-based mapping method. In order to achieve a high quality converted speech, these main problems have to be tackled.

A method of deploying LPC in voice conversion, extracting LPC Coefficients and LPC filter transfer function analysis has been discussed in [6]. Along with LPC analysis/encoding, which involves determining the voiced/unvoiced input speech, estimation of pitch period, filtering and transmitting the relevant parameters has been demonstrated by Bradbury in [7]. Also, the LPC source-filter model representation and LPC synthesis/decoding has been elaborated in [7].A basic understanding of incorporating GMM for voice transformation while still utilizing the LPC technique for extraction of coefficients has been illustrated by Gundersen in [8]. In [9]-[11], the GMM process has been elucidated with an approach that utilizes the concept of Dynamic Time Warping (DTW) and Expectation Maximization. However, the GMM approach deployed in [9]-[11] is based on a traditional approach that frequently results in over-smoothening of the source's speech. A modern approach to GMM which overcomes the over-smoothening drawback has been explained in [12]. We follow a similar approach and the corresponding execution is found to fetch good results and conclusions. In [13], a speaking-aid model for laryngectomees has been suggested. This paper proposes and analyses a system which deploys the model proposed in [13] with a modern GMM-based approach, thus not only overcoming the drawbacks of the traditional GMM-based approach for voice conversion but also improving the usability of the system for Laryngectomees.

## 2. IDEA OF THE PROPOSED SOLUTION

### 2.1 Phase 1

The source and destination voice will undergo LPC analysis, followed by the separation of glottal waveform. The vocal tract parameters will be subjected to GMM prediction.

### 2.2 Phase 2

The GMM based prediction is followed by the conversion rule based on GMM. The conversion rule has an input from the test speaker's voice, after being subjected to LPC analysis and separation of glottal waveform. The conversion rule will generate the output as the converted speech after LPC synthesis.

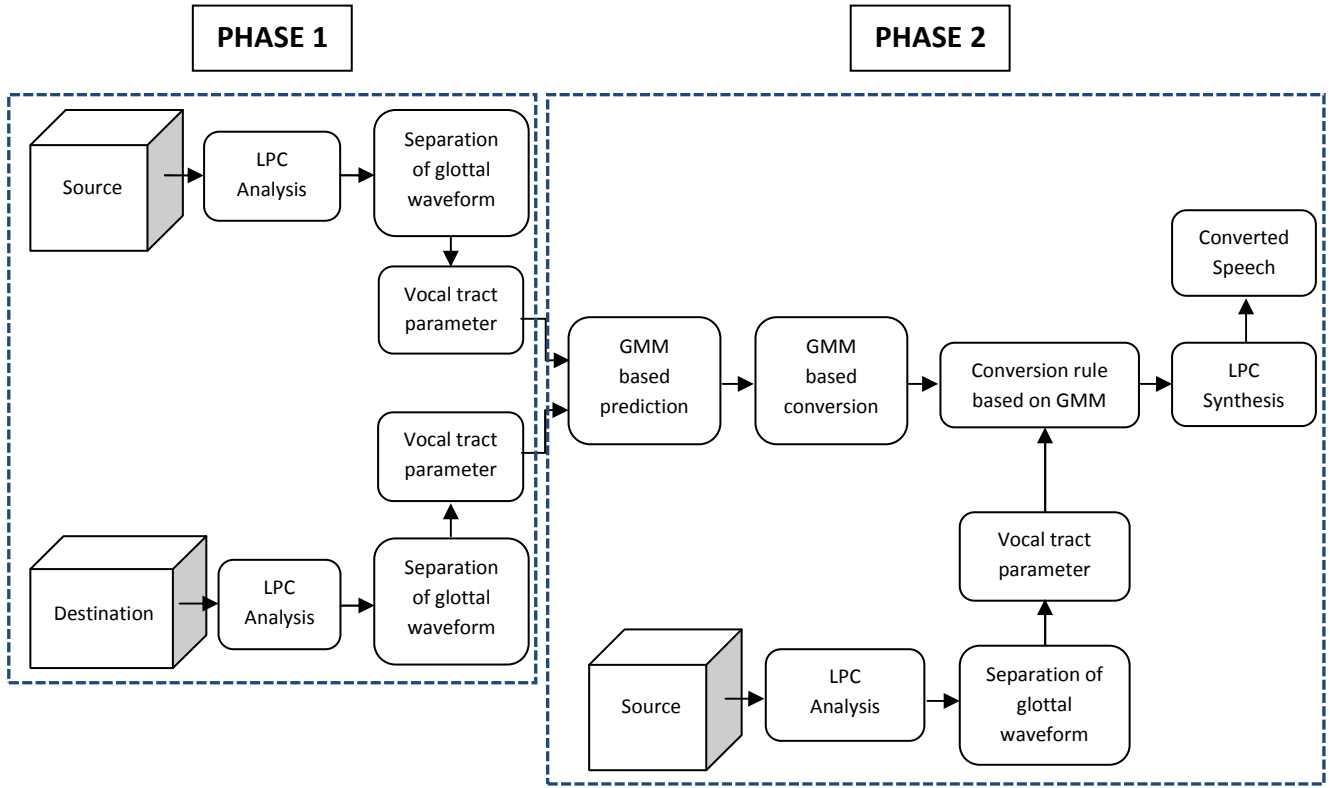

**Fig 1: Overview of the proposed system**

We hereby present voice morphing which involves separating out the excitation signal from the speech waveform of the speaker in order to convert vocal tract parameters in a precise manner. The remainder of this paper is organized as follows. Firstly, a review of voice conversion mechanism is given in this section, followed by the detailed description of technique proposed. Later on, there is an evaluation of the performance of the proposed system with the new technique. Finally, the overall applications, conclusions and results are presented at the end.

## 3.  OVERVIEW OF THE SYSTEM

The overview of our proposed voice conversion system is as shown in figure 1. The entire system comprises of two stages.

### 3.1.1  Training Stage

This includes segmentation of the source and target speaker voices in into equal frames of two pitch period lengths and then analyzing it is based upon a Linear Predictive Coding model. This is done in order to extract vocal features to be transformed [6][7].

### 3.1.2  Transforming Stage

This involves features transformation from source to target. In order to have a better alignment between source and the target features, there is an application of Dynamic Time Warping (DTW) in the preprocessing step.

## 3.2  Glottal waveform separation algorithm

According to the Linear Prediction algorithm, an effective speech production model for voiced speech is as shown in figure 2.

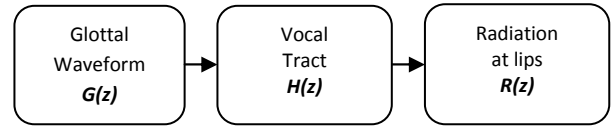

**Fig 2: Speech production model**

The terminologies used have been defined as follows:
S (z) is the acoustic speech waveform,
G (z) is the glottal waveform shaping,
V (z) is the models the vocal tract configuration and
R (z) is the radiation at the lips which can be modeled as an effect of the differential operator.

Now, G' (z) which is the glottal derivative, can be derived as the product of G(z) and R(z).

Hence,

$$S(z) = G'(z) \cdot V(z) = G(z) \cdot V(z) \cdot R(z) \qquad (1)$$

Given this assumption of the model, it was obvious that we can directly obtain the explicit representation of vocal tract by inverse filtering S (z) with G' (z).

Unfortunately, the premise does not hold valid as shown in [14]. Accurate and precise vocal tract parameters can only be obtained if and only if the glottal derivative effect is eliminated. Figure 3 shows the glottal derivative as well as the ideal glottal waveform in a pitch period.

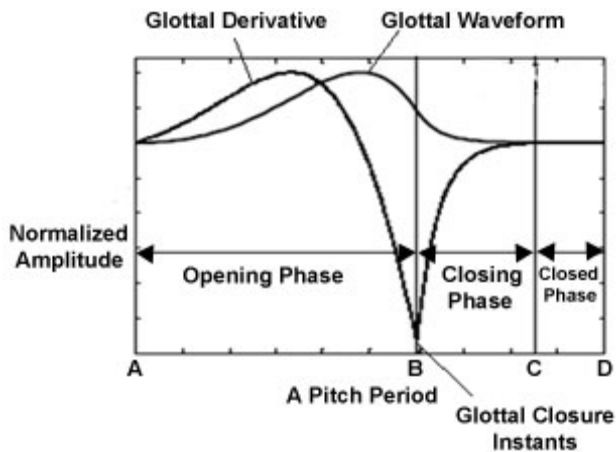

**Fig 3: Glottal derivative and glottal waveform**

It can be seen that the amplitude of the glottal waveform starting from the Glottal Closure Instants (GCI) to the end of the pitch period is decreasing monotonically, i.e., during the closing phase and the closed phase of the glottal waveform, the interaction with the vocal tract is decreased monotonically, where linear prediction coding (LPC) analysis can be performed in order to model the vocal tract almost exclusively since glottal contribution is minimum [1].

In order to achieve an effective personality change it is needed to change the glottal excitation characteristics of the source speaker to match exactly as that of the target, so a prediction rule has been trained on the aspects of the excitation signals of the target speaker. In the transforming stage, the source vocal tract features were extracted and modified based on the conversion rule from the training stage, meanwhile, converted excitation signals were obtained by predicting from the transformed vocal tract features based on the prediction rule [15][16]. Finally, a continuous waveform was obtained in the LPC synthesis model by synthesizing all these parameters.

## 4. IMPLEMENTATION STEPS

Transformation stage performance in voice conversion systems is generally evaluated using both objective and subjective measures. Both objective and subjective evaluations are essential to assess the performance of such systems. Objective evaluations are indicative of conversion performance and could be useful to compare different algorithms within a particular framework. However, objective measures on their own are not reliable, since they may not be directly correlated with human perception. As a result, a meaningful evaluation of voice conversion systems requires the use of subjective measures to perceptually evaluate their conversion outputs.

### 4.1 Database for System Implementation

A parallel database is sought wherein the source and the target speakers record a matching set of utterances. We narrowed down on the CMU ARCTIC databases [17], constructed at the Language Technologies Institute at Carnegie Mellon University, which consists of phonetically balanced, US English single speaker databases designed for unit selection speech synthesis research. Utterances recorded by six speakers are included in the database which is deployed for system evaluation. Each speaker has recorded a set of 1132 phonetically balanced utterances. The ARCTIC database includes utterances of SLT (US Female), CLB (US Female), RMS (US Male), JMK (Canadian Male), AWB (Scottish Male), KSP (Indian Male).

## 5. APPLICATION OF THE MODERN VOICE MORPHING APPROACH FOR LARYNGECTOMEES

Laryngectomy is a procedure of surgical removal of the larynx which is usually conducted on patients with laryngeal cancer or a similar problem. When the patient loses his larynx, he/she is called a laryngectomee, he/she would lose their original voice, and they will face difficulty interacting with other persons. An electrolarynx is an electronic speech aid that enables the user to communicate with other people as quickly as possible after the successful removal of the larynx. This device produces a tone that is transmitted over a membrane through the soft parts on the bottom of the chin into the mouth and throat cavity. In combination with clear articulation the tone is formed into speech. An electrolarynx is a popular method for enabling laryngectomees to speak without the vibration of the vocal cord due to the fact that it has several advantages as follows:

- Speaking with an electrolarynx can be learned quickly, which makes communication possible shortly after surgery.
- The electrolarynx can be used with an intraoral device while post-operative radiation therapy and if radiation treatment causes temporary loss of the esophageal voice.
- Using an electrolarynx is stressless, and not physically exhausting.
- An electrolarynx enables louder, faster and fluent speech right from the start.
- The device can be used in almost all situations, including during meals and whenever esophageal speech ceases or is not possible to use (i.e. in situations of stress, inflammation, radiation therapy, poor anatomical conditions etc.)
- The pitch of the electrolarynx can be individually adjusted.

Despite the distinct advantages of an electrolarynx, several disadvantages of an Electrolarynx exist:

- Not natural. The electrolarynx is a technical device. It can be defective, misplaced or can be rendered unusable by a dead battery.
- The sound of an electrolarynx is more noticeable than esophageal speech. A 'robotic voice' effect clearly persists in the sound of an electrolarynx. This results in unnaturalness of the artificial speech.
- The electrolarynx is a visual prosthesis.
- One hand is always occupied by the electrolarynx.
- Very clear articulation is requested.
- Accentuated speech can't be learned by all patients (depending on their motor skills and musicality).
- Leakage of sound source signals.

In order to make artificial speech sufficiently audible an electrolarynx needs to generate sufficiently loud sound source signals. Thus, the sound source signals themselves are noisy for other people around the laryngectomees especially in a quiet surrounding. Also, electrical imitation of any vocal fold vibration will cause a degradation of the speech quality.

A number of other alternatives to the electrolarynx have also been developed, however each was found to have its own set of disadvantages. An 'Artificial Larynx' apparatus proposed in [18], has indeed put an end to the search for a more lifelike and individualized voice by doing away with the robotic voice provided by a mechanical larynx, however it suffers from a 0.3-second delay between when the tongue and mouth move, and when the computer calculates the right word. This leaves the mouth and the voice out of sync, giving the user the appearance of being dubbed. An 'Electronic skin' proposed by Dae-Hyeong K. et al.in [19] is a wearable sensor that could help monitor health, amplify speech or control prosthetics, but suffers from the drawback of being expensive to manufacture. Another major downside of this system is that the continual shedding of skin cells means that the patch falls off after a few days.

A major point of concern is that the imitational speech cannot completely imitate the artificial speech uttered by laryngectomees. Thus, it is important to investigate how effective the system is, by using data obtained from laryngectomees. Although it is good for the user to get the auditory feedback from the current converted speech but higher quality voice morphing techniques that can work in real-time have not yet been devised. Thus, we propose a novel method to provide the signals detected by NAM microphone straight to the speaker as explicit auditory feedback. We hereby recommend a speaking-aid system using a modern voice conversion technique for laryngectomees to provide much natural speech communication [2].

## 5.1 Structure of the Proposed System

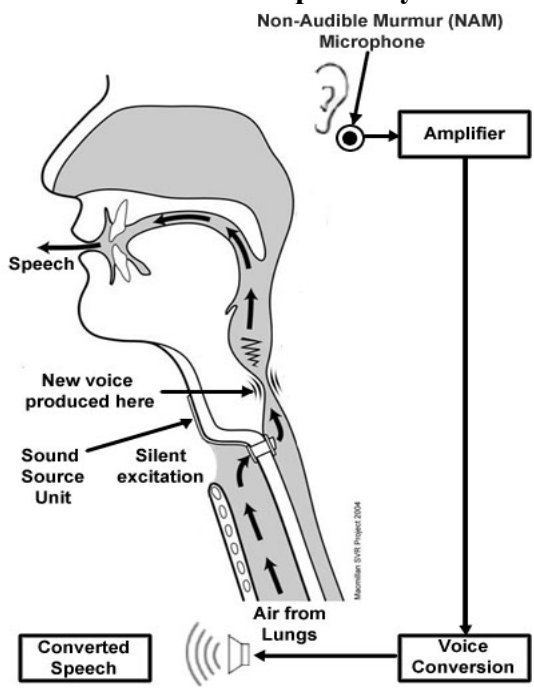

**Fig 4: Structure of the Proposed System**

As proposed in [13], figure 3 shows the speaking-aid system with a voice conversion technique for laryngectomees.

First, a user attaches a sound source unit under the lower jaw and articulates sound source signals. A distinct feature of the sound source signals is that the power is extremely small so that people around the user cannot hear it. To capture the small artificial speech, we use a Non-Audible Murmur (NAM) microphone powered with a high dynamic gain and a large range of operation. As the artificial speech detected with NAM microphone is still not the same as the natural speech, the captured data is converted into natural speech using our proposed voice conversion technique. Our system thus uses the modern GMM-based approach discussed in the earlier sections of this paper for accomplishing the voice conversion section depicted in the figure. Finally, the converted speech is presented as the voice of user [13].

## 6. PERFORMANCE ANALYSIS AND RESULTS

### 6.1 Subjective Evaluation

The ultimate objective of the voice conversion system is to transform an utterance of a speaker to sound as if spoken by the target speaker while maintaining the naturalness in speech. Hence, in order to evaluate the proposed system on these two scales, three types of subjective measures are generally used, as mentioned below:

- ABX Test
- MOS Test
- Similarity Test

#### 6.1.1 ABX Test

In order to check if the converted speech is perceived as the target speaker, ABX tests are most commonly used where participants listen to source (A), target (B) and transformed (X) utterances and are asked to determine whether A or B is closer to X in terms of speaker identity. A score of 100% indicates that all listeners find the transformed speech closer to the target.

#### 6.1.2 Mean Opinion Score and Similarity Test

The transformed speech is also generally evaluated in terms of naturalness and intelligibility by Mean Opinion Score (MOS) tests. In this test, the participants are asked to rank the transformed speech in terms of its quality and/or intelligibility. Listeners evaluate the speech quality of the converted voices using a 5-point scale, where:

- 5: excellent
- 4: good
- 3: fair
- 2: poor
- 1: bad

This is similar to the similarity test, but the major difference lies in the fact that we concentrate on the speaker characteristics in the similarity test and intelligibility in the MOS test.

The MOS test is generally preferred over the Similarity test as we are more interested in the intelligibility, hence the MOS test has been formed and the results have been tabulated in Table 1.

**Table 1. MOS Training Table**

| No. of Gaussians | No. of test speaker sample for training | Quality |
|---|---|---|
| 1 | 2 | 2.9 |
| 5 | 2 | 3 |
| 10 | 2 | 3.5 |
| 1 | 8 | 3 |
| 5 | 8 | 4 |
| 10 | 8 | 4.3 |

Here the training is done with two cases:

- Using two samples each of source and target
- Using eight samples each of source and target

After training we have introduced a test speaker which will be an input to the training and will exactly reproduce the voice of the target speaker. Hereafter we have done three tests:

- Signal to noise distortion
- Average spectral distortion
- Varying number of Gaussians

The implementation thus follows.

Upon varying the number of Gaussians as (1, 5, 10) and number of test speaker samples as (2, 8), we have assigned a rating as per the observations.

## 6.2 Objective Evaluation

Objective evaluations are indicative of conversion performance and could be useful to compare different algorithms within a particular framework. We have calculated the signal to noise ratio of the transformed speech and also the average spectral distortion. We have tested with two test speaker sample training and eight test speaker sample training and compared the results.

**Table 2. Average distortion for each sample**

| Test speaker sample number | Average spectral distortion |
|---|---|
| 1 | 2.940637 |
| 2 | 1.359387 |
| 4 | 3.050596 |
| 5 | 1.370122 |
| 6 | 1.836406 |
| 1 | 2.940637 |

### 6.2.1 Signal to Noise Ratio

A Signal to Noise Ratio (SNR) of less than 1.1 suggests a very poor quality of signal. Using the proposed method, we are getting a good quality signal having an SNR within the range of 2.5 to 3.6.We have picked the test speaker samples from test folder 1-6 and calculated the signal to noise ratio for both 2 test speaker samples and 8 test speaker samples training.

**Table 3. SNR for 2 and 8 test speaker samples**

| Test speaker sample number | 2 test speaker training sample SNR (dB) | 8 test speaker training sample SNR (dB) |
|---|---|---|
| 1 | 3.65 | 3.6505 |
| 2 | 2.5578 | 2.5608 |
| 4 | 3.0162 | 3.0100 |
| 5 | 3.4803 | 3.4867 |
| 6 | 3.2115 | 3.2100 |
| 1 | 3.5851 | 3.5851 |

### 6.2.2 Average spectral distortion

Average spectral distortion is the objective evaluation of differences between two speech signals. We have taken 6 test speakers and individually found out the difference between the test signal and the target signal for 8 sample training (source + target).

**Table 4. Total objective evaluation**

| No. of test speaker sample for training | No. of Gaussians | Time for execution (seconds) | SNR (dB) | Average spectral distortion |
|---|---|---|---|---|
| 2 | 1 | 10 | 3.6550 | 2.946 |
| 2 | 3 | 20 | 3.6585 | 2.936 |
| 2 | 5 | 40 | 3.6582 | 2.943 |
| 2 | 10 | 180 | 3.6539 | 2.936 |
| 8 | 1 | 30 | 3.6591 | 2.946 |
| 8 | 3 | 45 | 3.6563 | 2.944 |
| 8 | 5 | 70 | 3.6567 | 2.943 |
| 8 | 10 | 300 | 3.6512 | 2.945 |

Thus, we observe that for 2 test speaker samples used for training, the signal to noise ratio was maximum for number of Gaussians lying between 3 to 5 whereas in the case of 8 samples being used for training, the signal to noise was maximum when the number of Gaussians was 1. For traditional systems the signal to noise ratio was around about 2.1 to 2.6 whereas for this method it is well above 3 as depicted by the table above.

## 7. CONCLUSION AND FUTURE SCOPE

This project presents a novel method which is based on the technique of the separation of glottal waveforms and the prediction of the transformed residuals for precise voice conversion. Performance analysis and results show that not only are the transformed vocal tract parameters matching the target one better, but also are the target personalities preserved.

Although the enhancements described in this paper give a substantial improvement, there is still distortion remained which makes the audio quality depressive and the future work will therefore focus on it.

During the practical implementation of the system, in certain cases, some critical problems may arise. The difficulty of acquiring the auditory feedback would cause a negative spiral of instability of the articulation, the degradation of the converted speech quality and more instability of the articulation. This would thus result in a repetitive degradation and instability process. Steps to overcome this issue need to be worked upon in future course of time.

## 8. ACKNOWLEDGMENTS

Our sincere thanks to Prof. Rakesh Chadha, Biology Department, St. Xavier's College, for contributing towards the development of the proposed system by providing us with information regarding Laryngectomy and the newer solutions being deployed (apart from an electrolarynx) to enable smooth communication for laryngectomees.

The authors would also like to thank Prof. J. H. Nirmal, Department of Electronics Engineering, K. J. Somaiya College of Engineering, for his prompt guidance regarding the project.

## AUTHOR BIOGRAPHIES

**Aman Chadha** (M'2008) was born in Mumbai (M.H.) in India on November 22, 1990. He is currently pursuing his graduate studies in Electrical and Computer Engineering at the University of Wisconsin-Madison, USA. He completed his B.E. in Electronics and Telecommunication Engineering from the University of Mumbai in 2012. His special fields of interest include Signal and Image Processing, Computer Vision (particularly, Pattern Recognition) and Processor Microarchitecture. He has 9 papers in International Conferences and Journals to his credit. He is a member of the IETE, IACSIT and ISTE.

**Bharatraaj Savardekar** (M'2008) was born in Mumbai (M.H.) in India on January 24, 1990. He completed his B.E. in Electronics Engineering from the University of Mumbai in 2012. His fields of interest include Audio Processing and Human-Computer Interaction.

**Jay Padhya** (M'2008) was born in Mumbai (M.H.) in India on May 16, 1990. He completed his B.E. in Electronics Engineering from the University of Mumbai in 2012. His fields of interest include Audio Processing and Human-Computer Interaction.